\documentclass{ifacconf}

\usepackage{graphicx}      
\usepackage{natbib}        
\usepackage{url}
\usepackage{amsmath}
\usepackage{amsfonts}
\usepackage{amssymb}
\begin{document}
\begin{frontmatter}

\title{Characterizing the Predictive Accuracy of Dynamic Mode Decomposition for Data-Driven Control}

\thanks[footnoteinfo]{Corresponding author.}

\author[First]{Qiugang Lu} 
\author[First]{Sungho Shin} 
\author[First]{Victor M. Zavala\thanksref{footnoteinfo}} 

\address[First]{Department of Chemical and Biological Engineering, University of Wisconsin-Madison, 1415 Engineering Dr., Madison, WI 53706, USA (e-mail: qlu62@wisc.edu; sungho.shin@wisc.edu; victor.zavala@wisc.edu).}

\begin{abstract}                
Dynamic mode decomposition (DMD) is a versatile approach that enables the construction of low-order models from data. Controller design tasks based on such models require estimates and guarantees on predictive accuracy. In this work, we provide a theoretical analysis of DMD model errors that reveals impacts of model order and data availability. The analysis also establishes conditions under which DMD models can be made asymptotically exact. We numerically validate our theoretical results using a 2D diffusion system. 
\end{abstract}


\end{frontmatter}

\section{Introduction}
Dynamic mode decomposition (DMD) is a versatile approach that enables the construction of low-order models from data \citep{schmid2010dynamic,tu2014dynamic}. DMD uses singular value decomposition (SVD) to extract modes that describe dominant patterns in the data and temporal frequencies associated with those modes \citep{zhang2019online}. Interesting connections between DMD and Koopman operator theory have been recently established by \cite{mezic2013analysis}. DMD has been applied to video processing \citep{kutz2016dynamicVideo}, fluid dynamics \citep{tissot2014model}, and finance \citep{mann2016dynamic}.

A key feature of DMD is that it builds low-order models directly from data and thus contrasts with traditional model reduction approaches such as balanced truncation \citep{moore1981principal,troltzsch2009pod} (which seek to reduce an {\em existing} full-order model). Moreover, compared with other reduction techniques such as 
proper orthogonal decomposition (POD), DMD extracts dominant modes that are \text{both} spatially and temporally critical for the system \citep{zhang2017evaluating,berkooz1993proper}. Thus, DMD  enables modeling and control of high-dimensional systems (e.g., spatiotemporal) for which a full order representation might be difficult or impossible to construct \citep{proctor2016dynamic,korda2018linear}. A control application based on DMD models in wind farms has been recently reported by \citep{annoni2016wind}. 

Despite the wide applicability of DMD, the characterization of its predictive accuracy remains an open problem. \cite{korda2018convergence} established  asymptotic convergence of extended DMD models to Koopman operators with respect to the sample size. An empirical analysis of model accuracy of DMD is conducted by \cite{duke2012error} using synthetic waveforms that resemble instability structures in shear flows.  \cite{zhang2017evaluating} proposes a criterion to evaluate the accuracy of each DMD mode against the corresponding Koopman modes. More recently, \cite{lu2019predictive} develops an upper bound on the predictive error of DMD; here, the error is defined as the difference between DMD predicted states and the true states (obtained with a full-order model). We emphasize that these works characterize predictive accuracy for autonomous systems (with no control). To the best of our knowledge, characterizations of the predictive accuracy of DMD models with control (often called DMDc) have not been reported.  Characterizing the predictive accuracy of such models is essential for controller design (e.g., establishing stability and robustness properties). We highlight that characterizations of predictive accuracy of DMDc differ from those of classical model reduction methods such as balanced truncation (for which extensive literature exist) because DMDc builds low-order models directly from data. 

In this work, we present a theoretical analysis of the predictive accuracy of DMDc models. We derive an explicit error bound that reveals the effect of model order and the number of data samples. Our analysis also establishes conditions under which the error vanishes. These insights indicate that DMDc provides a coherent approach for data-driven control. We provide a case study for a 2D diffusion system to illustrate the developments. The paper is organized as follows:  In Section \ref{DMDc} we summarize the DMDc algorithm. In Section \ref{Section3} we analyze the prediction error of DMDc and establish theoretical properties. In Section \ref{Section4} we present a 2D diffusion system to verify the theoretical results. Conclusions and a preview of future work are presented in Section \ref{Section5}. 

\section{DMD with Control (DMDc)} \label{DMDc}

DMDc uses data snapshots of state measurements and inputs to construct a low-order representation of the full-order model: 
\begin{equation}
\mathbf{x}_{k+1} = \mathbb{A} \mathbf{x}_{k} + \mathbb{B} \mathbf{u}_{k},  \label{TrueLinearModel}
\end{equation}
where $\mathbb{A}\in \mathbb{R}^{n\times n}$ and $\mathbb{B}\in \mathbb{R}^{n\times q}$, are the system matrices. We are interested in systems with high state dimension (order) $n\gg 1$. Symbol $\mathbf{x}_{k}$ denotes the state vector and $\mathbf{u}_{k}$ is the control input. In our setting, the full-order model \eqref{TrueLinearModel} represents the ground truth. Given a sample of size $m$, we stack the state-input observations in the data matrix:
\begin{align}
\mathbf{X}&=\left[\mathbf{
	x_{1}~x_{2}~\cdots~x_{m-1}}
\right]\nonumber\\
 \mathbf{Y}&=\left[\mathbf{
	x_{2}~x_{3}~\cdots~x_{m}}
\right] \nonumber \\
\mathbf{\Upsilon}&=\left[\mathbf{
	u_{1}~u_{2}~\cdots~u_{m-1}}
\right]. \label{Data}
\end{align}
Model \eqref{TrueLinearModel} can be expressed in matrix-data form as:
\begin{align}
\mathbf{Y} &= \mathbb{A} \mathbf{X} + \mathbb{B} \mathbf{\Upsilon}\nonumber\\
& = \mathbf{\Theta} \mathbf{\Omega}, \label{TrueDataFormat}
\end{align}
where $ \mathbf{\Theta}:=[\mathbb{A}~\mathbb{B}]$ and $\mathbf{\Omega}:=[\mathbf{X}^{T}~\mathbf{\Upsilon}^{T}]^{T}$. In DMDc, the system matrices embedded in $ \mathbf{\Theta}$ are estimated by minimizing the Frobenius norm of the residuals $\|\mathbf{Y}-\mathbf{\Theta\Omega}\|_{F}$. This can be done by using a truncated SVD of $\mathbf{\Omega}$ \citep{proctor2016dynamic}, 
\begin{equation}
\hat{\mathbf{\Theta}}=\mathbf{Y}\mathbf{\Omega}_s^{\dagger}=\mathbf{Y}\hat{\mathbf{V}}_{s}\hat{\mathbf{\Sigma}}_{s}^{-1}\hat{\mathbf{U}}_{s}^{T}, \label{FullOrderTruncatedSVD}
\end{equation}
where $s$ is truncation order; $\mathbf{\Omega}_s= \hat{\mathbf{U}}_{s}\hat{\mathbf{\Sigma}}_{s}\hat{\mathbf{V}}_{s}^{T}$ is the rank-$s$ approximation of $\Omega$; $(\cdot)^\dag$ is the psuedoinverse of the argument. This least-squares solution induces an error between $\hat{\mathbf{\Theta}}$ and $\mathbf{\Theta}$. Partitioning $\hat{\mathbf{\Theta}}$ yields the estimated system matrices:
\begin{align}
\hat{\mathbf{A}}=\mathbf{Y}\hat{\mathbf{V}}_{s}\hat{\mathbf{\Sigma}}_{s}^{-1}\mathbf{\hat{U}}_{1,s}^{T}, ~ \hat{\mathbf{B}}=\mathbf{Y}\hat{\mathbf{V}}_{s}\hat{\mathbf{\Sigma}}_{s}^{-1}\hat{\mathbf{U}}_{2,s}^{T},  \label{FullOrderEstimates}
\end{align}
with $\hat{\mathbf{U}}_{1,s}\in \mathbb{R}^{m\times s}$, $\hat{\mathbf{U}}_{2,s}\in \mathbb{R}^{q\times s}$, and  $\hat{\mathbf{U}}^{T}_{s}=[\hat{\mathbf{U}}^{T}_{1,s}~\hat{\mathbf{U}}^{T}_{2,s}]$.   These matrices give the approximated representation:
\begin{equation}
\mathbf{x}_{k+1} \approx \mathbf{\hat{A}} \mathbf{x}_{k} + 
\mathbf{\hat{B}} \mathbf{u}_{k},  \label{Ahat}
\end{equation}
We note that $\hat{\mathbf{A}}$ and $\hat{\mathbf{B}}$ are of the same dimension as $\mathbb{A}$ and $\mathbb{B}$. Consequently, when $n$ is large, storing $\hat{\mathbf{A}}$ and $\hat{\mathbf{B}}$ and performing computations with them can cause tractability issues.  This motivates the use of low-order model representations. Direct use of $\hat{\mathbf{U}}_s$ to reduce the model does not provide a suitable basis because this spans the joint space of states and inputs (while we are only interested in reducing the state space). A common approach (discussed by \cite{kutz2016dynamic}) to handle this is to use a second truncated SVD on $\mathbf{Y}$:
\begin{equation}
\mathbf{Y}=\mathbf{U}_{r}\mathbf{\Sigma}_{r}\mathbf{V}_{r}^{T}, \label{SVD_Y}
\end{equation}
with truncation order $r \le s \ll n$. The columns of $\mathbf{U}_{r}$ are used to obtain the low-order model:
\begin{equation}
\tilde{\mathbf{x}}_{k+1} = \tilde{\mathbf{A}} \tilde{\mathbf{x}}_{k} + \tilde{\mathbf{B}} \mathbf{u}_{k}, \label{ROM}
\end{equation}
where $\tilde{\mathbf{x}}_{k}\in \mathbb{R}^{r}$, $\tilde{\mathbf{A}}\in\mathbb{R}^{r\times r}$, $\tilde{\mathbf{B}}\in\mathbb{R}^{r\times q}$, with expressions
\begin{align}
\tilde{\mathbf{A}} &= \mathbf{U}_{r}^{T}\hat{\mathbf{A}}\mathbf{U}_{r}=\mathbf{U}_{r}^{T}\mathbf{Y}\hat{\mathbf{V}}_{s}\hat{\mathbf{\Sigma}}_{s}^{-1}\mathbf{\hat{U}}_{1,s}^{T}\mathbf{U}_{r}, \label{ReducedOrderEstimateAtilde}\\  \tilde{\mathbf{B}} &= \mathbf{U}_{r}^{T}\hat{\mathbf{B}}=\mathbf{U}_{r}^{T}\mathbf{Y}\hat{\mathbf{V}}_{s}\hat{\mathbf{\Sigma}}_{s}^{-1}\hat{\mathbf{U}}_{2,s}^{T}. \label{ReducedOrderEstimateBtilde}
\end{align}
The true state can be approximated as $\mathbf{x}_{k}\approx \hat{\mathbf{x}}_{k}=\mathbf{U}_{r}\tilde{\mathbf{x}}_{k}$ where $\hat{\mathbf{x}}_{k}$ is the reconstructed state. Leading eigenvalues and eigenvectors $(\mathbf{\Lambda},\mathbf{\Phi})$ of the matrix $\hat{\mathbf{A}}$ are computed through the eigendecomposition $\tilde{\mathbf{A}}=\mathbf{W\Lambda W}^{-1}$, with 
\begin{equation}
\mathbf{\Phi}=\mathbf{U}_{r}\mathbf{W}. \label{DMDEigenvectors}
\end{equation}
Instead of using exact DMD (as in \cite{tu2014dynamic}), we use  
\eqref{DMDEigenvectors} to compute the eigenvectors of $\hat{\mathbf{A}}$ (this is known as the projected DMD) and has been explored in \cite{schmid2010dynamic}. An attractive feature associated with DMDc is that computations rely entirely on the data matrices and do not involve operations in the high-dimensional space ($\hat{\mathbf{\Theta}}$ is not explicitly needed to obtain a low-order model). 

We define the error induced by the DMDc model as: 
\begin{equation}
\mathbf{e}_{k}=\mathbf{x}_{k}-\hat{\mathbf{x}}_{k}. \label{ReconstructionError}
\end{equation}
This is a reconstruction error if $k\le m$ and a predictive error if $k>m$. Since  $\mathbf{x}_{k}$ and $\hat{\mathbf{x}}_{k}$  are dynamic, the error $\mathbf{e}_{k}$ has dynamics as well. In this work, we seek to characterize the dynamics of this error and identify critical factors that affect the error amplitude. 

\begin{rm}
As shown in \cite{kutz2016dynamic}, under mild conditions, DMDc is a special case of an identification technique known as eigensystem realization algorithm (ERA).  A key difference of DMDc with traditional subspace identification methods such as  MOSEP, N4SID and CVA is that these first identify a (possibly low-order) state sequence from the block Hankel data matrix (stacking delay-embedded data in \eqref{Data}), followed by least-squares to estimate system matrices. The resultant state sequence conveys similar compressed information as the reduced-order state vector $\tilde{\mathbf{x}}$ in \eqref{ROM} from DMDc. A formal study of the relationship between DMDc and subspace identification methods is an important topic for future work.
\end{rm}

\section{Predictive Accuracy of DMDc} \label{Section3}
In this section we characterize the predictive accuracy of DMDc. We derive an upper bound on the prediction error and this reveals important factors that can be tuned to improve model accuracy. We also establish conditions under which the model can be asymptotically exact. 

\subsection{Preliminaries}
We introduce assumptions and technical results that are necessary for the subsequent analysis. 

\textit{Assumption 1.} \label{Asm1}
For model \eqref{TrueLinearModel}, let $\{\lambda_1,\ldots,\lambda_{n}\}$ be the eigenvalues of $\mathbb{A}$. We assume that $\mathbb{A}$ is Hurwitz with spectral radius:
\begin{equation}
\rho(\mathbb{A}):=\max_{i=1,\ldots,n} |\lambda_{i}| < 1. \label{Hurwitz}
\end{equation}
It will be shown that the dynamics of the error $\mathbf{e}_{k}$ depend on the eigenvalues of $\mathbb{A}$. The system matrix $\mathbb{A}$ being Hurwitz ensures that the error $\mathbf{e}_{k}$ decays asymptotically over time (in the absence of inputs). 

\begin{lem}\label{Lem1}
For any matrix $\mathbf{A}\in \mathbb{R}^{n\times n}$ with spectral radius $\rho(\mathbf{A})<1$, there exists a constant $\bar{\rho}\in (\rho(\mathbf{A}),1)$ and a constant $M\ge1$ such that, for each nonnegative integer $k$, 
\begin{equation}
\|\mathbf{A}^{k}\| \le M \bar{\rho}^{k}. \label{Lemma1}
\end{equation}
\end{lem}

\begin{pf}
	The proof follows from Gelfand's formula; thus is omitted. \hfill $\blacksquare$
\end{pf}


If the matrix $\mathbb{A}$ is symmetric (e.g., when the system model \eqref{TrueLinearModel} is obtained by discretizing PDEs)  we have that  $\|\mathbb{A}^{k}\|$ is bounded by powers of its spectral radius $\|\mathbb{A}^{k}\| \le \rho(\mathbb{A})^k$. The following corollary can be established from Lemma \ref{Lem1}. 
\begin{cor}\label{Cor1}
Consider square matrices $\mathbf{A},\mathbf{R}\in \mathbb{R}^{n\times n}$, where $\mathbf{A}$ is Hurwitz. For some constant $M>0$ and $\bar{\rho} \in (\rho(\mathbf{A}),1)$,
\begin{equation}
\|\mathbf{R}\mathbf{A}^{k}\| \le M\bar{\rho}^{k}. 
\end{equation}
\end{cor}
The proof to Corollary \ref{Cor1} follows from the submultiplicative property of matrix norms. 

\subsection{Error Dynamics for DMDc Model}
We first examine the error incurred from the estimation of $\hat{\mathbf{\Theta}}$ in \eqref{FullOrderTruncatedSVD}. 

\begin{thm} \label{Thm1}
 Consider the true model \eqref{TrueLinearModel} and the least-squares estimate \eqref{FullOrderEstimates} obtained via the truncated SVD \eqref{FullOrderTruncatedSVD} and define also the true solution as $\mathbf{\Theta}$. The estimation error is given by:
\begin{equation}
\| \hat{\mathbf{\Theta}}-\mathbf{\Theta}\| = \|\mathbf{\Theta}( \mathbf{I}-\hat{\mathbf{U}}_{s}\hat{\mathbf{U}}_{s}^{T})\| = \varepsilon_{s}, \label{FullOrderErrorBound}
\end{equation} 
where $\varepsilon_{s} $ is a constant that decreases as the truncation order $s$ increases. Furthermore, the estimation errors for $\hat{\mathbf{A}}$ and $\hat{\mathbf{B}}$ are: 
\begin{equation}
\|\mathbb{A}-\hat{\mathbf{A}}\| = \varepsilon^{A}_{s}, ~\|\mathbb{B}-\hat{\mathbf{B}}\| = \varepsilon^{B}_{s}, \label{HatBound}
\end{equation}
where $\varepsilon^{A}_{s}$ and $\varepsilon^{B}_{s}$ are constants decreasing as $s$ increases. 
\end{thm}

\begin{pf}
The full-order SVD of $\mathbf{\Omega}$ can be expressed as: 
\begin{equation}
\mathbf{\Omega}=\left[\hat{\mathbf{U}}_{s}~\hat{\mathbf{U}}_{e} \right]
\left[
\begin{array}{cc}
\hat{\mathbf{\Sigma}}_{s} & \mathbf{0} \\
\mathbf{0} & \hat{\mathbf{\Sigma}}_e
\end{array}
\right]
\left[
\begin{array}{c}
\hat{\mathbf{V}}_{s}^{T} \\
\hat{\mathbf{V}}_{e}^{T}
\end{array}
\right], \label{FullSVD}
\end{equation}
where $\hat{\mathbf{U}}_{e}$ and $\hat{\mathbf{V}}_{e}$ represent collections of the remaining left and right singular vectors of $\mathbf{\Omega}$, respectively. Substituting \eqref{TrueDataFormat} into \eqref{FullOrderTruncatedSVD} and applying \eqref{FullSVD} we obtain:
\begin{align}
\hat{\mathbf{\Theta}} &= \mathbf{\Theta} \mathbf{\Omega}\mathbf{\Omega}_s^{\dagger}\nonumber\\
&=\mathbf{\Theta}\left[\hat{\mathbf{U}}_{s}~\hat{\mathbf{U}}_{e} \right]
\left[
\begin{array}{cc}
\hat{\mathbf{\Sigma}}_{s} & \mathbf{0} \nonumber\\
\mathbf{0} & \hat{\mathbf{\Sigma}}_e
\end{array}
\right]
\left[
\begin{array}{c}
\hat{\mathbf{V}}_{s}^{T}\nonumber \\
\hat{\mathbf{V}}_{e}^{T}
\end{array}
\right]\hat{\mathbf{V}}_{s}\hat{\mathbf{\Sigma}}_{s}^{-1}\hat{\mathbf{U}}_{s}^{T}\\
&=\mathbf{\Theta}\hat{\mathbf{U}}_{s}\hat{\mathbf{U}}_{s}^{T}.
\end{align}
We can thus obtain the error between the estimated $\hat{\mathbf{\Theta}}$ and the true matrix $\mathbf{\Theta}$ as $\mathbf{\Theta}-\hat{\mathbf{\Theta}}=\mathbf{\Theta}( \mathbf{I}-\hat{\mathbf{U}}_{s}\hat{\mathbf{U}}_{s}^{T})$. We thus have that \eqref{FullOrderErrorBound} follows. Note that $\mathbf{\Theta}( \mathbf{I}-\hat{\mathbf{U}}_{s}\hat{\mathbf{U}}_{s}^{T})$ can be interpreted as the projection of the row space of $\mathbf{\Theta}$ onto the orthogonal complement of the space spanned by the columns of $\hat{\mathbf{U}}_{s}$. We thus have that, as the truncation order $s$ increases,  $\|\mathbf{\Theta}( \mathbf{I}-\hat{\mathbf{U}}_{s}\hat{\mathbf{U}}_{s}^{T})\| $ decreases. The error expressions for $\hat{\mathbf{A}}$ and $\hat{\mathbf{B}}$ follow directly. \hfill $\blacksquare$
\end{pf}

As shown in Theorem \ref{Thm1}, $\varepsilon_{s}$, $\varepsilon_{s}^{A}$ and $\varepsilon_{s}^{B}$ are directly related to the truncation order $s$ in the SVD of $\mathbf{\Omega}$.The selection of $s$ is determined by the distribution of singular values of $\mathbf{\Omega}$, which has a close relationship with the quality of the input signal $\mathbf{\Upsilon}$. Moreover, analogous to system identification theory, when there is noise contaminating the data, a 
larger sample size can give a smaller covariance of the estimates. Therefore, with properly designed input signals, a reasonably higher truncation order $s$ and larger sample size $m$ can be used to reduce the estimation error in the presence of noise. We also note that an excessively large order $s$ may render the estimates overly sensitive to noise. 

The following is our main result. 

\begin{thm}\label{Theorem2}
Consider system \eqref{TrueLinearModel} and its low-order approximation \eqref{ROM} obtained from DMDc. Under Assumption 1, the prediction error $\mathbf{e}_{k}$ (for $k>m$) can be bounded as:
\begin{align}
&\|\mathbf{e}_{k}\| \le  M\bar{\rho}^{k-m} \|\mathbf{e}_{m}\|+M(k-m)\bar{\rho}^{k-1-m}(M_{s,m}+M_{r,m})  \nonumber \\ &\quad  \cdot\|\mathbf{x}_{m}\| +M(\varepsilon_{s}^{B}+\varepsilon_{r}^{B})\sum\nolimits_{i=0}^{k-1-m}  \bar{\rho}^{k-1-m-i}\|\mathbf{u}_{i+m}\| +  \nonumber \\ & \quad M (M_{s,m}+ M_{r,m})\sum\nolimits_{i=0}^{k-2-m}(i+1)\bar{\rho}^{i}\|\mathbb{B}\mathbf{u}_{k-2-i}\|, \label{Thm2}
\end{align} 
where $\bar{\rho}\in(\rho(\mathbb{A}),1)$, $M>0$ is a constant such that $\|\mathbf{\Phi}\mathbf{\Lambda}^{k}\mathbf{\Phi}^{-1}_{lf}\| \le M \bar{\rho}^{k}$, $\forall k>0$, and positive constants $M_{s,m}$, $M_{r,m}$, $\varepsilon_{r}^B$, and $\varepsilon_{s}^B$ decrease as the SVD truncation orders $r$, $s$, and the sample size $m$ increase. 
\end{thm}
  
\begin{pf}
From the definition of $\mathbf{e}_{k}$ and noting that $\tilde{\mathbf{x}}_{k}=\mathbf{U}_{r}^{T}\hat{\mathbf{x}}_{k}$,
we have
\begin{align*}
\mathbf{e}_{k} &= \mathbf{x}_{k} - \mathbf{\hat{x}}_{k} = \mathbf{x}_{k}-\mathbf{U}_{r} \tilde{\mathbf{x}}_{k} \\
&= \mathbb{A}\mathbf{x}_{k-1} + \mathbb{B} \mathbf{u}_{k-1} -\mathbf{U}_{r}\tilde{\mathbf{A}}\tilde{\mathbf{x}}_{k-1} - \mathbf{U}_{r}\tilde{\mathbf{B}} \mathbf{u}_{k-1} \\
&= \mathbb{A}(\mathbf{x}_{k-1}-\hat{\mathbf{x}}_{k-1}) + \mathbb{A} \hat{\mathbf{x}}_{k-1} + \mathbb{B} \mathbf{u}_{k-1} \\
& \quad  -\mathbf{U}_{r}\tilde{\mathbf{A}}\tilde{\mathbf{x}}_{k-1} - \mathbf{U}_{r}\tilde{\mathbf{B}} \mathbf{u}_{k-1} \\
&=\mathbb{A}\mathbf{e}_{k-1} + (\mathbb{A}-\mathbf{U}_{r}\tilde{\mathbf{A}}\mathbf{U}_{r}^{T}) \hat{\mathbf{x}}_{k-1} + (\mathbb{B}-\mathbf{U}_{r}\tilde{\mathbf{B}})\mathbf{u}_{k-1} \\
&=\mathbb{A} \mathbf{e}_{k-1} + (\mathbb{A}-\mathbf{U}_{r}\tilde{\mathbf{A}}\mathbf{U}_{r}^{T}) (\mathbf{x}_{k-1}-\mathbf{e}_{k-1}) \\ 
&\quad +(\mathbb{B}-\mathbf{U}_{r}\mathbf{\tilde{B}}) \mathbf{u}_{k-1} \\
&=\mathbf{U}_{r}\tilde{\mathbf{A}}\mathbf{U}_{r}^{T} \mathbf{e}_{k-1} +  (\mathbb{A}-\mathbf{U}_{r}\tilde{\mathbf{A}}\mathbf{U}_{r}^{T}) \mathbf{x}_{k-1} \\
& \quad + (\mathbb{B}-\mathbf{U}_{r}\mathbf{\tilde{B}}) \mathbf{u}_{k-1}  \\
&= \mathbf{\Phi} \mathbf{\Lambda} \mathbf{\Phi}^{-1}_{lf} \mathbf{e}_{k-1} + (\mathbb{A}-\mathbf{\Phi \Lambda \Phi}^{-1})\mathbf{x}_{k-1}  \\
& \quad + (\mathbb{B}-\mathbf{U}_{r}\mathbf{\tilde{B}}) \mathbf{u}_{k-1}.
\end{align*}
Here, the last equality follows from \eqref{DMDEigenvectors} and $\mathbf{\Phi}^{-1}_{lf}:=\mathbf{W}^{-1}\mathbf{U}_{r}^{T}$ is defined as the left inverse of $\mathbf{\Phi}$. Iterating the above equation over $k$ yields 
\begin{align}
\mathbf{e}_{k} &= \mathbf{\Phi \Lambda}^{k-m} \mathbf{\Phi}^{-1}_{lf} \mathbf{e}_{m} + \sum\nolimits_{i=0}^{k-1-m} \mathbf{\Phi} \mathbf{\Lambda}^{i} \mathbf{\Phi}^{-1}_{lf} \mathbf{\Psi}_{k-1-i}, 
\label{ek}
\end{align}
where $\mathbf{\Psi}_{i}=(\mathbb{A}-\mathbf{\Phi} \mathbf{\Lambda} \mathbf{\Phi}^{-1}_{lf}) \mathbf{x}_{i}+(\mathbb{B}-\mathbf{U}_{r}\tilde{\mathbf{B}})\mathbf{u}_{i}$. When the sample size $m$ is large, we can safely assume $\hat{\mathbf{A}}$ to be Hurwitz. This implies that $\mathbf{\Phi} \mathbf{\Lambda}^{k} \mathbf{\Phi}^{-1}_{lf}$ is stable with bounded norm for all $k\ge 0$. Using the definition of $M$ we can establish that $\|\mathbf{e}_{k}\|$ in  \eqref{ek} satisfies
\begin{align}
\|\mathbf{e}_{k}\| &\le M\bar{\rho}^{k-m} \|\mathbf{e}_{m}\| + M\sum\nolimits_{i=m}^{k-1} \bar{\rho}^{k-1-i} \|\mathbf{\Psi}_{i}\|.
\label{ek_norm}
\end{align}
Now let us focus on $\mathbf{\Psi}_{i}$ with $m\le i\le k-1$; this can be expressed as 
\begin{align}
\mathbf{\Psi}_{i} &=(\mathbb{A}-\hat{\mathbf{A}})\mathbf{x}_{i}+ (\hat{\mathbf{A}}-\mathbf{\Phi} \mathbf{\Lambda} \mathbf{\Phi}^{-1}_{lf}) \mathbf{x}_{i}+(\mathbb{B}-\hat{\mathbf{B}})\mathbf{u}_{i} \nonumber \\
&\qquad    +(\hat{\mathbf{B}}-\mathbf{U}_{r}\tilde{\mathbf{B}})\mathbf{u}_{i} \nonumber\\
&=(\mathbb{A}-\hat{\mathbf{A}})\mathbb{A}^{i-m}\mathbf{x}_{m}+\sum\nolimits_{j=0}^{i-1-m}(\mathbb{A}-\hat{\mathbf{A}})\mathbb{A}^{j}\mathbb{B}\mathbf{u}_{i-1-j} + \nonumber \\
& \quad  (\hat{\mathbf{A}}-\mathbf{\Phi} \mathbf{\Lambda} \mathbf{\Phi}^{-1}_{lf}) \mathbb{A}^{i-m}\mathbf{x}_{m} + \sum\nolimits_{j=0}^{i-1-m}(\hat{\mathbf{A}}-\mathbf{\Phi} \mathbf{\Lambda} \mathbf{\Phi}^{-1}_{lf}) \nonumber \\
& \quad ~\cdot\mathbb{A}^{j}\mathbb{B}\mathbf{u}_{i-1-j} + (\mathbb{B}-\hat{\mathbf{B}})\mathbf{u}_{i} + (\mathbf{I}-\mathbf{U}_{r}\mathbf{U}_{r}^{T})\hat{\mathbf{B}}\mathbf{u}_{i} \label{Psi}.
\end{align}
As shown in \cite{korda2018convergence}, the estimated DMD modes converge to the true Koopman modes when the sample size $m$ tends to infinity. Moreover, it is shown that larger SVD truncation orders increase the accuracy of DMD, provided that the inverse of singular values in relevant computations does not cause numerical issue or stability problem. Thus, we can establish that: 
\begin{equation}
\|\hat{\mathbf{A}}-\mathbf{\Phi} \mathbf{\Lambda} \mathbf{\Phi}^{-1}_{lf}\| \le c_{r,m}, \label{cmr}
\end{equation}
where $c_{r,m}>0$ is a constant decreasing as $r$ or $m$ increases. The results in Lemma \ref{Lem1}, Corollary \ref{Cor1}, and Theorem \ref{Thm1}, yield: 
\begin{align}
&\|(\mathbb{A}-\hat{\mathbf{A}})\mathbb{A}^{i}\| \le M_{s,m}\bar{\rho}^{i},~\|(\hat{\mathbf{A}}-\mathbf{\Phi} \mathbf{\Lambda} \mathbf{\Phi}^{-1}_{lf})\mathbb{A}^{i}\| \le M_{r,m} \bar{\rho}^{i}, \nonumber \\
&\|(\mathbf{I}-\mathbf{U}_{r}\mathbf{U}_{r}^{T})\hat{\mathbf{B}}\| \le \varepsilon_{r}^{B}, \label{Constants}
\end{align}
where $M_{s,m}>0$, $M_{r,m}>0$  are constants decreasing as $s$, $r$, or $m$ increases and the constant $\varepsilon_{r}^{B}>0$ decreases as $r$ increases. Combining \eqref{Constants} and \eqref{Psi} we obtain
\begin{align}
&\sum_{i=m}^{k-1} \bar{\rho}^{k-1-i} \|\mathbf{\Psi}_{i}\|
\le  (k-m)\bar{\rho}^{k-1-m}(M_{s,m}+M_{r,m})\|\mathbf{x}_m\| \nonumber \\
& + (\varepsilon_{s}^{B}+\varepsilon_{r}^{B})  \sum_{i=0}^{k-1-m} \bar{\rho}^{k-1-m-i} \|\mathbf{u}_{i+m}\|+(M_{s,m}+M_{r,m})\cdot \nonumber \\
& 
\sum_{i=0}^{k-1-m}\sum_{j=0}^{i-1}\bar{\rho}^{k-1-m-i+j} \|\mathbb{B}\mathbf{u}_{i+m-1-j}\|. \label{Psi_norm}
\end{align}
With change of variables, the result in \eqref{Thm2} follows by substituting \eqref{Psi_norm} into \eqref{ek_norm}. \hfill $\blacksquare$

\end{pf}

Theorem \ref{Thm1} provides a characterization of the error dynamics of DMDc models. The result reveals the role of the sample size $m$ and SVD truncation orders $s$ and $r$. According to \eqref{Thm2}, a larger sample size and higher SVD truncation orders reduce the error (as expected) and asymptotic convergence is obtained. However, similar to the error analysis of balanced truncation presented in \cite{moore1981principal}, the predictive errors of DMDc depend on the amplitude of the input $\mathbf{u}_{k}$. This effect is not analyzed here and is left as a topic of future work. We also highlight that the error bound obtained in Theorem \ref{Theorem2} is not tight, and alternative bounding approaches are also an important topic of future work. 

We now show that as $k\to\infty$, the prediction error is upper bounded if the amplitude of the input signal is bounded. 
\begin{cor}\label{Cor2}
For the prediction error bound obtained from Theorem \ref{Theorem2} with finite sample size $m$ and fixed truncation orders $s,r \le n$, assume that the input is bounded $\|\mathbf{u}_{k}\|\le \bar{u}$. As $k\to\infty$, the prediction error is upper bounded by 
\begin{equation}
\lim_{k\to\infty}\|\mathbf{e}_{k}\| \le \frac{M\bar{u}}{1-\bar{\rho}}(\varepsilon_{s}^{B}+\varepsilon_{r}^{B})+\frac{M\|\mathbb{B}\|\bar{u}}{(1-\bar{\rho})^2}(M_{s,m}+M_{r,m}).
\end{equation}	
\end{cor}
\begin{pf}
The first two terms in \eqref{Thm2} vanish as $k\to\infty$ since $|\bar{\rho}|<1$. For the last two terms, we can establish the result by using the fact that $\lim_{l\to\infty} \sum_{k=0}^{l}z^{k}=1/(1-z)$ and $\lim_{l\to\infty} \sum_{k=0}^{l}(k+1)z^{k}=1/(1-z)^2$, for all $|z|<1$.  \hfill $\blacksquare$
\end{pf}

\begin{figure}
	\begin{center}
		\includegraphics[width=6.5cm]{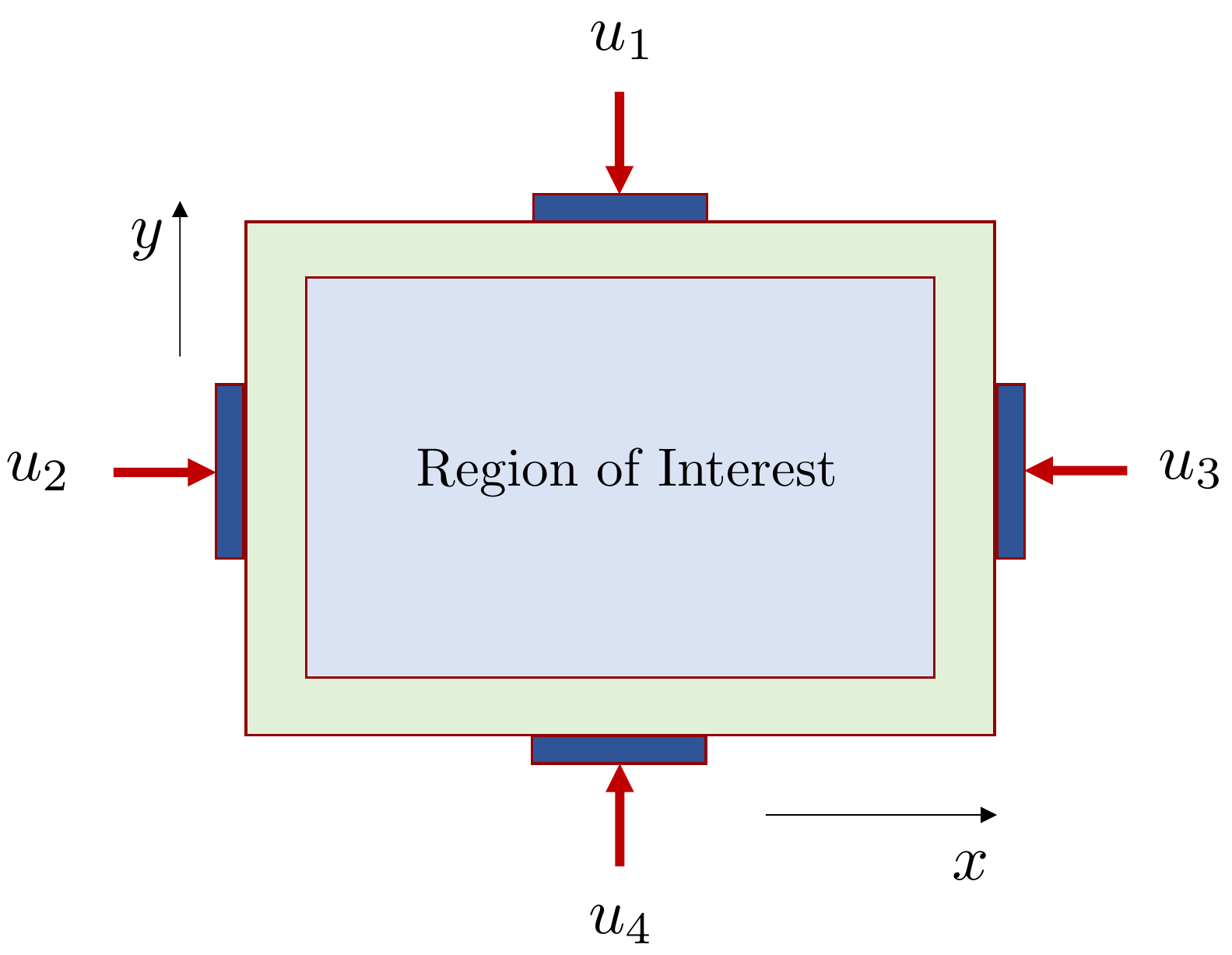}    
		\caption{Sketch of 2D diffusion system used for validating DMDc errors. } 
		\label{2DDiffusionSchematics}
	\end{center}
\end{figure}

\begin{figure*}[tbh]
\centering
	\includegraphics[width=0.9\textwidth]{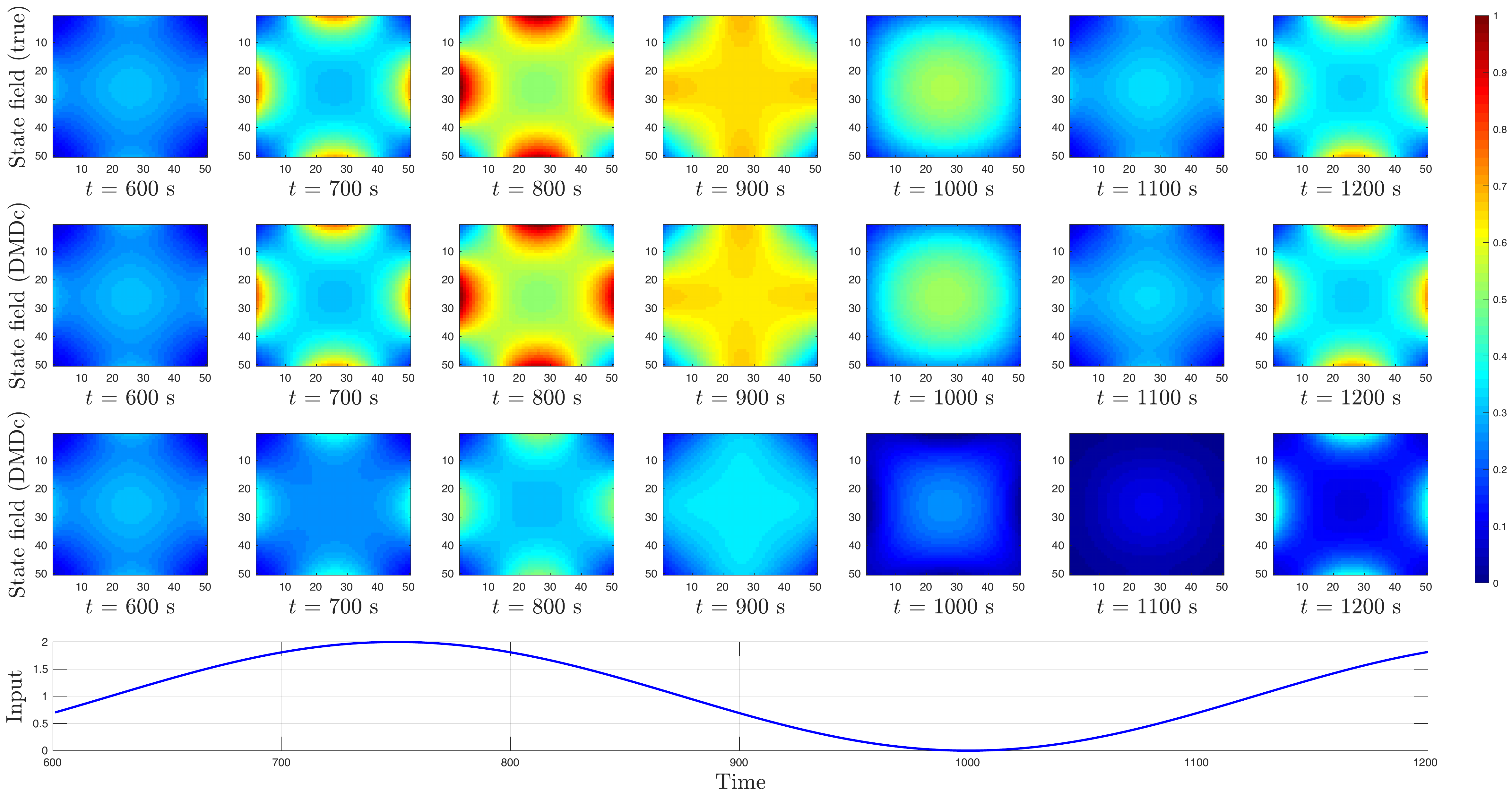}    
	\caption{Predictive accuracy of DMDc for 2D state field. Row 1: True state field; Row 2: Predicted field with DMDc using sample size $m=600$; Row 3: Predicted field with DMDc using sample size $m=350$; Row 4: Input signal used for data generation.} 
	\label{TrueDMDCompare}
\end{figure*}

\begin{rm}
Note that the theorems above demonstrate the pointwise bounds of the prediction error at a specific time instant from DMDc. As a comparison, for the balanced truncation, the corresponding Hankel-norm error bound is in terms of the $H_\infty$ norm of the error dynamic system, i.e., it measures the peak energy (or $L_2$) gain of the error signal over all types of inputs, across time from zero to infinity, instead of at a specific time instant.

\end{rm}

\section{Case Study} \label{Section4}
We use a 2D heat diffusion system to verify our proposed error bound for the predictive accuracy of DMDc.  In this experiment, our aim is to imitate a heat diffusion process along a 2D domain with heat sources located at four edges, as shown in Fig. \ref{2DDiffusionSchematics}. Each of the four heat source (inputs) $u_1$, $u_2$, $u_3$ and $u_4$ in Fig. \ref{2DDiffusionSchematics} spans a width along the edge, similar to the heat radiator on the wall of a room. Defining $\xi(a,b,t)$ as the temperature value at location $(a,b)$ at time $t$, the diffusion equation reads as
\begin{align}
&\frac{\partial \xi}{\partial t}=\alpha \left(\frac{\partial^2 \xi}{\partial a^2} + \frac{\partial^2 \xi}{\partial b^2}\right) + f(a,b), \label{2DDiffusion} \\
&(a,b)\in (0,L_a) \times (0,L_b),~t\in(0,T], \nonumber
\end{align} 
where $\alpha$ is the diffusion coefficient, $f(a,b)$ is the source at location $(a,b)$, $L_a$, $L_b$ are the spatial length along $a$ and $b$ directions, and $T$ is the simulation duration. $\xi(L_a,0,t) = \xi_a, ~ \xi(0,L_b,t) = \xi_b$ are the boundary conditions and $\xi(a,b,0) = I(a,b)$ is the initial condition. 

We use backward Euler scheme to simulate the evolution of \eqref{2DDiffusion} and we use a central finite difference scheme in the spatial domain. We set $L_a=L_b=40$ and discretize the 2D space into a $71\times 71$ mesh with $\Delta a=\Delta b = 0.5714$. We also set $\Delta t=1 s$ and $\alpha=0.45$.  To construct our ground truth model, we identify the high-dimensional system \eqref{TrueLinearModel} from the 2D diffusion equation \eqref{2DDiffusion}, where $\mathbb{A}$ in \eqref{TrueLinearModel} is required to be Hurwitz according to Assumption 1.   To eliminate the adverse effects from boundary conditions, we shrink our interested region to an inner domain (a $50\times 50$ mesh) of the 2D space, as highlighted in the blue area in Fig. \ref{2DDiffusionSchematics}. Each heat source spans 21 discrete spatial locations and so in total there are $q=84$ heating points mounted evenly on four edges to heat up the 2D board. We used pseudo random binary signals to excite the diffusion process and collected data over an horizon of $T=20,000$ seconds. The data was used to identify the state-space matrix $\mathbb{A}\in\mathbb{R}^{2500\times 2500}$ and $\mathbb{B}\in \mathbb{R}^{2500\times 84}$ in \eqref{TrueLinearModel} and we verified matrix $\mathbb{A}$ is stable.

To examine the predictive accuracy of DMDc, we simulated the high-dimensional model \eqref{TrueLinearModel} under sinusoidal inputs for all heat actuators. For simplicity, it is assumed that all inputs are identical. The amplitude and frequency of the sinusoidal signal are 2 and $0.02$ Hz, respectively.  The simulation time was set to $T=1,200$ seconds and this covers more than two periods of the sinusoidal input. We first used $m=600$ samples of data to construct the low-order model \eqref{ROM} with DMDc. Here, the SVD truncation orders are $s=26$ and $r=17$ and these were set based on the singular values of $\mathbf{\Omega}$ and $\mathbf{Y}$, respectively. The next 600 samples of data were used to assess the predictive performance of the DMDc model. Fig. \ref{TrueDMDCompare} illustrates that the DMDc model is accurate (despite this being of low order) but accuracy strongly depends on data availability.  

\begin{figure}[tbh]
	\begin{center}
		\includegraphics[width=0.97\columnwidth]{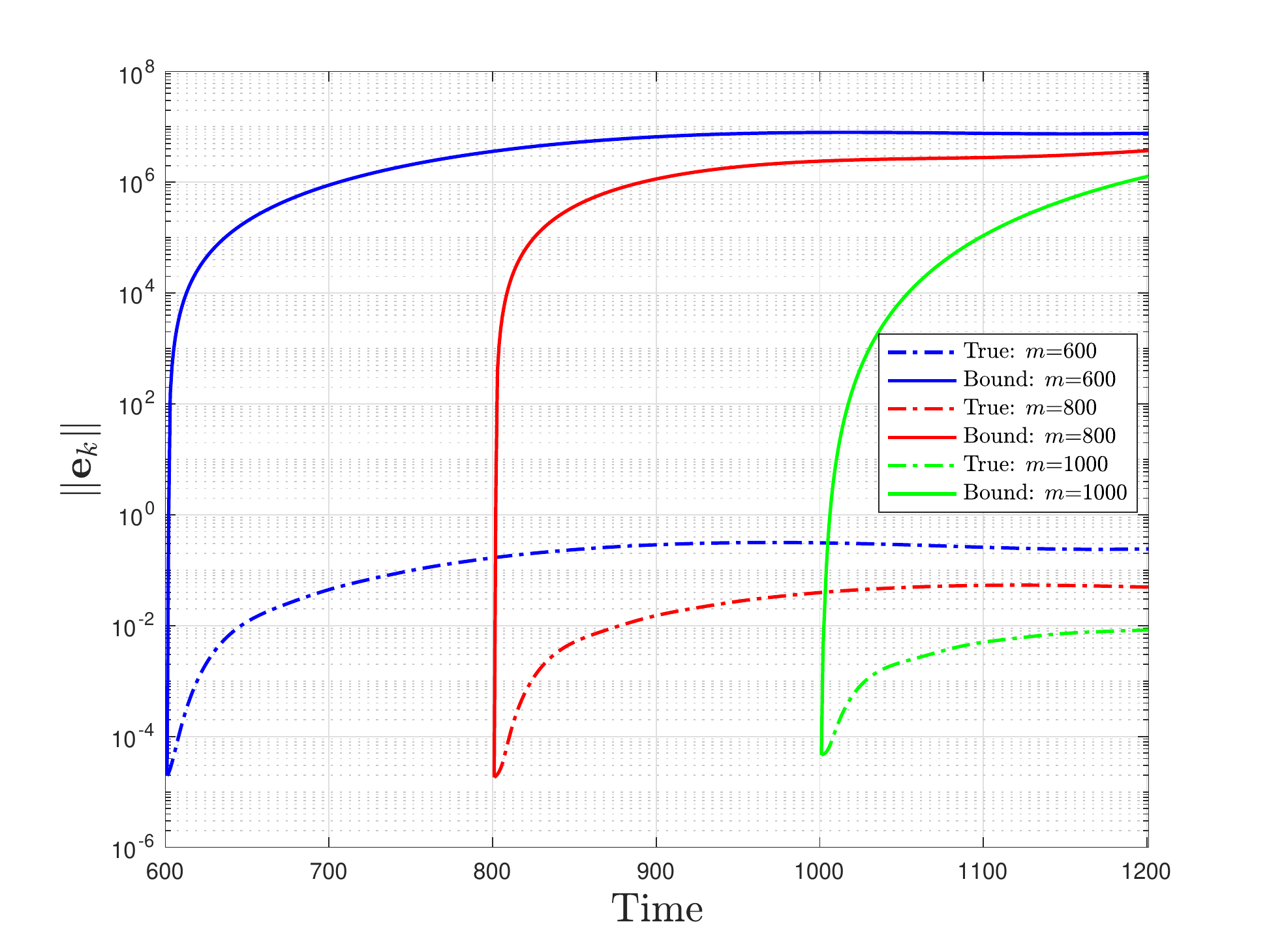}    
		\caption{Error bounds and actual errors for DMDc under different sample sizes $m$.} 
		\label{ErrorBoundSampleSize}
	\end{center}
\end{figure}

We now evaluate the impact of data size and order on the predictive accuracy. To this end, we separately test the predictive errors against varying values of $m$, $s$, and $r$.  For $m$, we select values of 600, 800, and 1000, while keeping $s=26$ and $r=17$. The prediction error bound was computed using \eqref{Thm2} and the true  error was computed using \eqref{ROM}. Fig. \ref{ErrorBoundSampleSize} shows that, as the sample size increases, the error bound and actual error decrease. We also see that the error stabilizes as one moves forward in the horizon (consistent with the error dynamics predicted). 

\begin{figure}[tbh]
	\begin{center}
		\includegraphics[width=0.97\columnwidth]{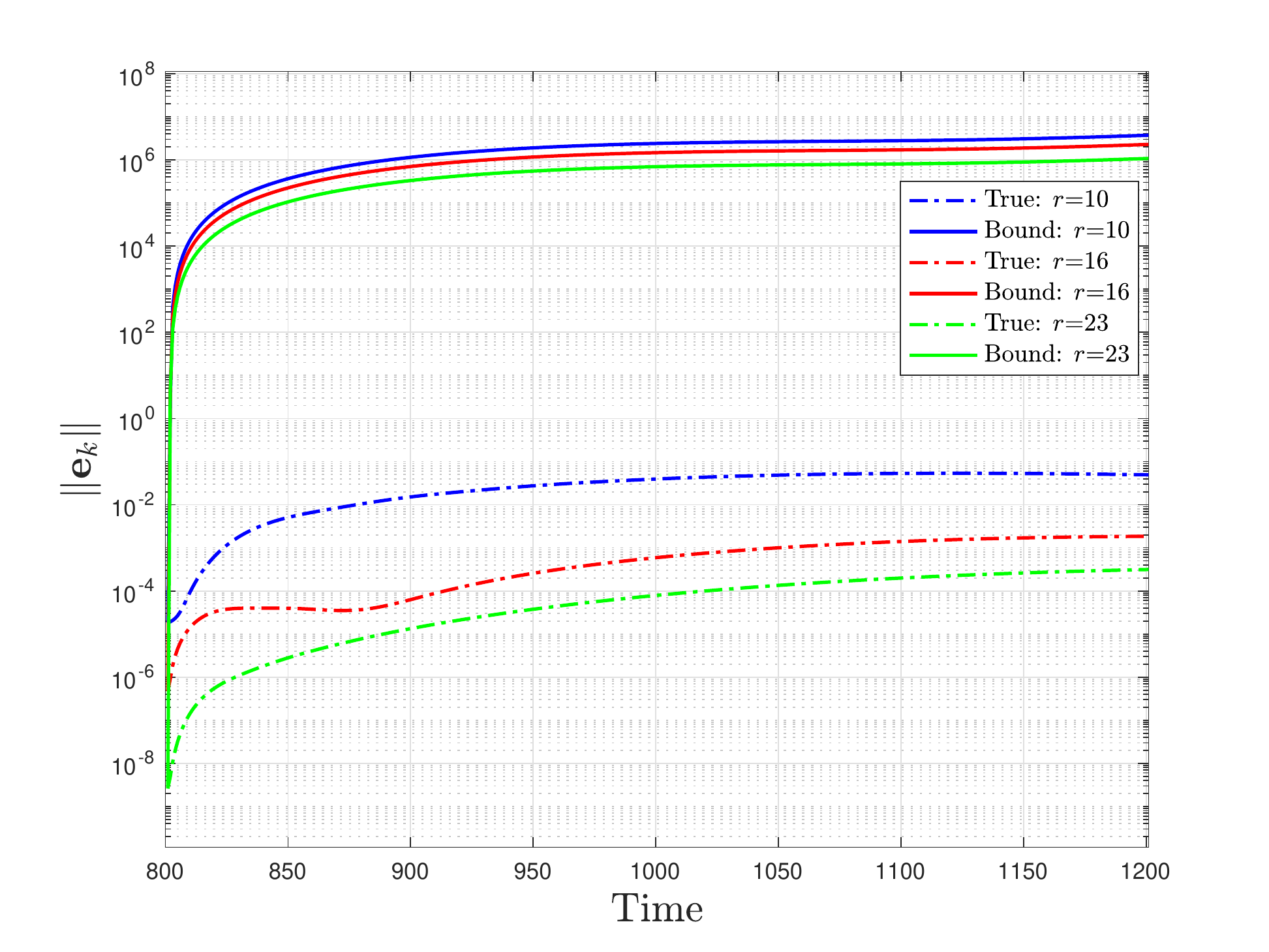}    
		\caption{Error bounds and actual errors for DMDc under different truncation orders $r$.} 
		\label{ErrorBoundr}
	\end{center}
\end{figure}

\begin{figure}[tbh]
	\begin{center}
		\includegraphics[width=0.97\columnwidth]{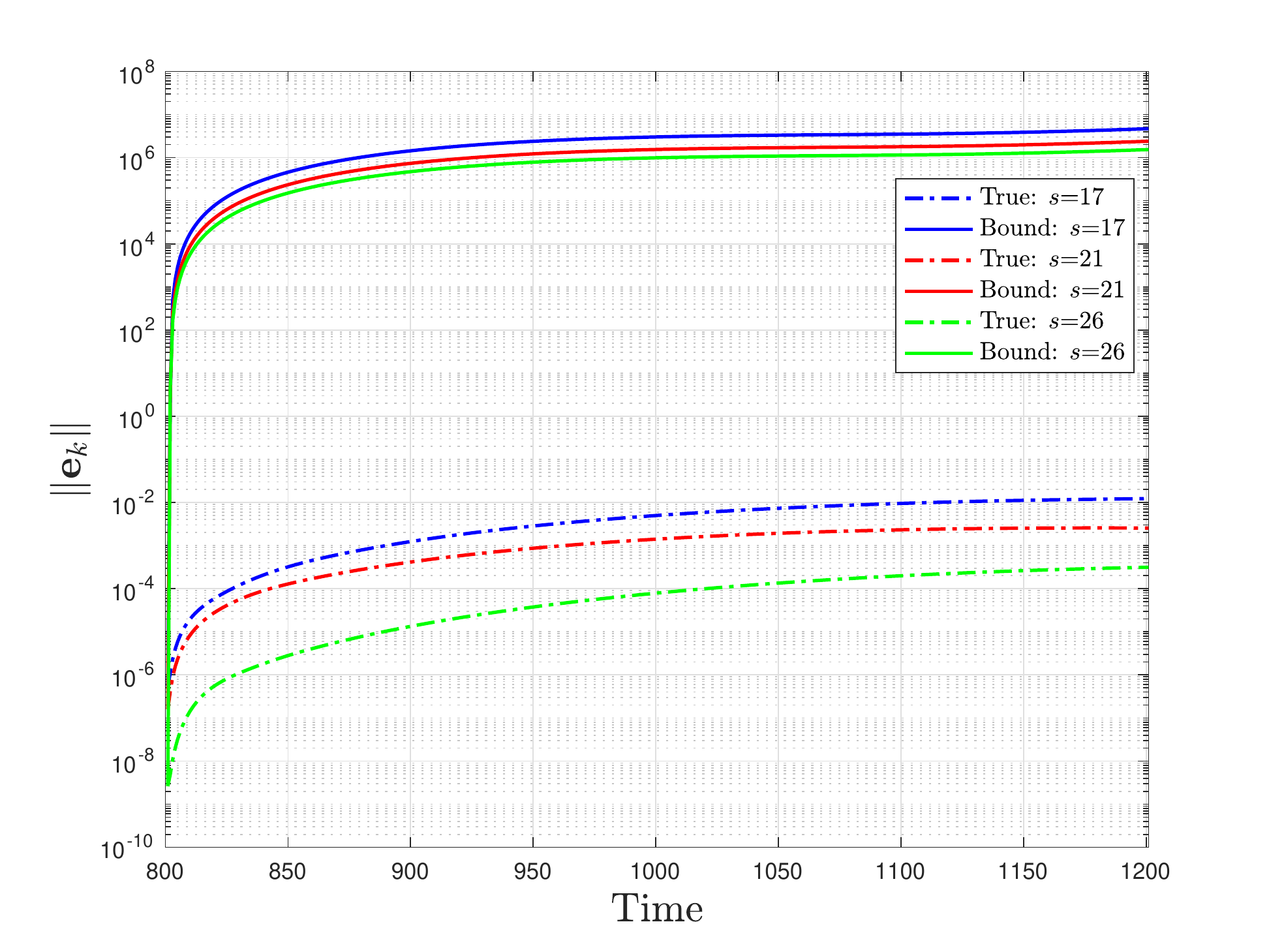}    
		\caption{Error bounds and actual errors for DMDc under different truncation orders $s$.} 
		\label{ErrorBounds}
	\end{center}
\end{figure}

\begin{figure}[tbh]
	\begin{center}
		\includegraphics[width=0.97\columnwidth]{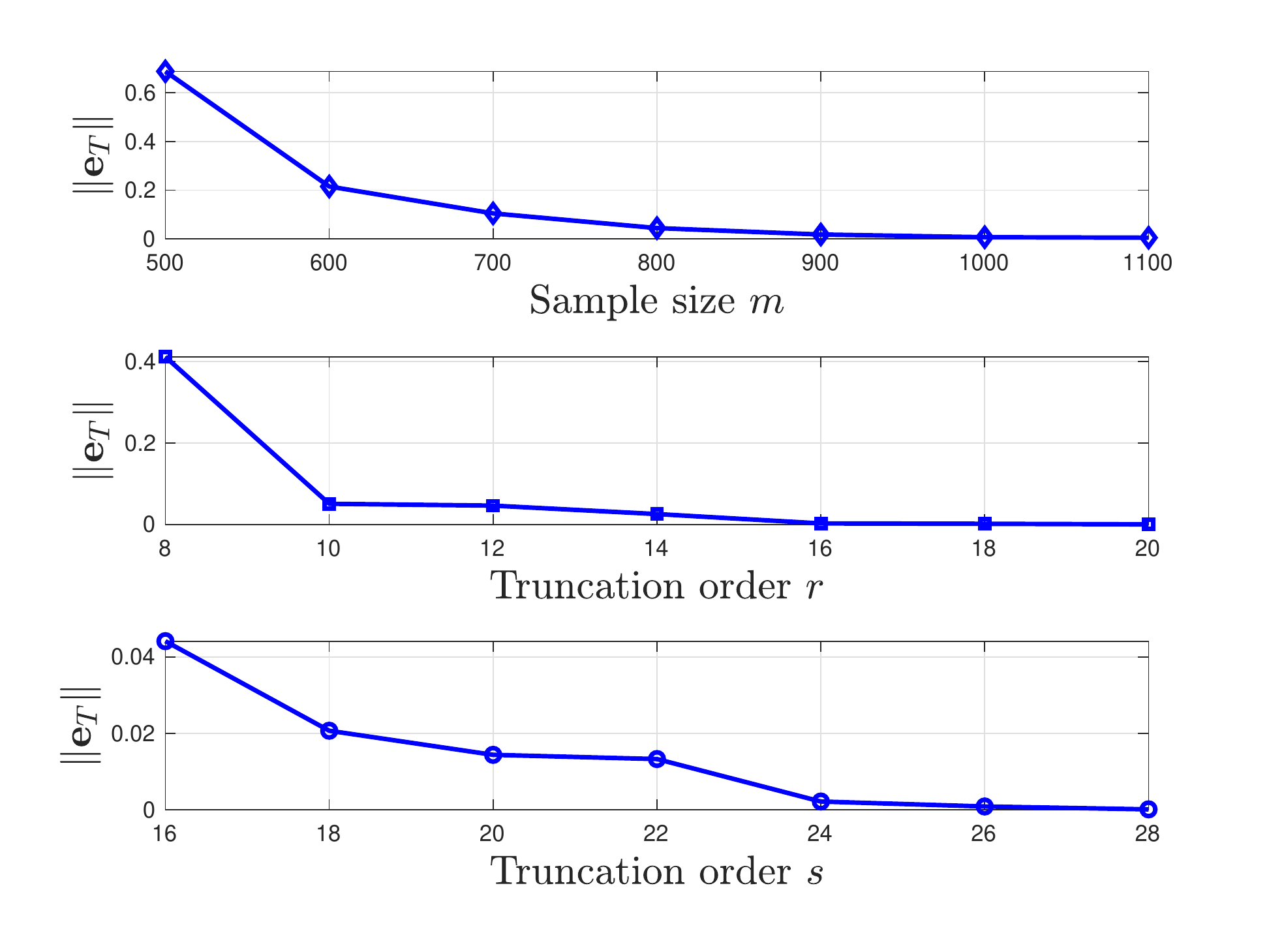}    
		\caption{Predicted terminal error $\|\mathbf{e}_{T}\|$ under different truncation orders and sample sizes.} 
		\label{ErrorPlot}
	\end{center}
\end{figure}

We next obtained the errors using truncation orders $r=10,16,23$ (with $s=26$ and $m=800$). Fig. \ref{ErrorBoundr} confirms the trend predicted by our theoretical bound (larger truncation orders lead to smaller errors).  For the last scenario, we fix the sample size to $m=800$ and alter the values of $s$. Here, we found that the contributed errors from both $r$ (being fixed) and $s$ change over different values of $s$ and this gives misleading results because we are only interested in examining the effects of $s$. This is due to the fact that the objective of the second SVD, determined by $r$ in \eqref{SVD_Y}, is to estimate $\hat{\mathbf{A}}$ from the first SVD. If $r$ is fixed but $s$ varies (and thus $\hat{\mathbf{A}}$ varies), the gap between $r$ and $s$ will change, causing the contributed errors from $r$ to be different. Hence, here we set $r=s-3$ (recall that $r\le s$ in the DMDc procedure) so that the order $r$ changes with $s$. Fig. \ref{ErrorBounds} shows error trajectories for  $s=17,21,26$. We can see that increasing the truncation order $s$ gives lower error bounds and actual error, confirming our theoretical results. Finally, we examined the norm of the error  at the end of the prediction horizon $\|\mathbf{e}_{T}\|$ for different sample sizes and orders. Fig. \ref{ErrorPlot} shows that the error decreases as the truncation orders $s$ and $r$ and sample size $m$ increase and we observe asymptotic convergence. This confirms that DMDc provides a coherent modeling approach for data-driven control.

%

\section{Conclusions and Future Work} \label{Section5}

In this work we characterize the prediction errors of DMDc. Our analysis highlights the effect of  SVD truncation orders and data samples on the error dynamics and provides conditions under which the error can be driven to zero. This analysis provides consistent guarantees for DMDc models that are desirable from a control stand-point. As part of future work, we will investigate alternative strategies to provide tighter error bounds. Moreover, we will seek to design model-predictive control based on ROMs from DMDc to control high-dimensional systems. The relationship between DMDc with subspace identification methods will also be explored.

\begin{ack}
The authors acknowledge support from the members of the TWCCC consortium.
\end{ack}

\bibliography{ifacconf}             
                                                   







\end{document}